\documentstyle[12pt]{article}
\def\simgt{\rlap{\lower 3.5 pt \hbox{$\mathchar \sim$}} \raise 1pt
  \hbox {$>$}}
\begin{document}
\thispagestyle{empty}

\hfill\vbox{\hbox{\bf DESY 94-212}
            \hbox{November 1994}
                                }
\vspace{1.in}
\begin{center}
{\large\bf Squark Production at the Tevatron} \\
\vspace{0.5in}
W.~Beenakker, R.~H\"opker, M.~Spira and P.~M.~Zerwas \\
\vspace{0.5in}
Deutsches Elektronen-Synchrotron DESY, D-22603 Hamburg, FRG \\
\end{center}
\vspace{2in}

\begin{center}
ABSTRACT \\
\end{center}
We have determined the QCD corrections to the production of
squark-antisquark pairs in $p\bar p$ collisions at the
Tevatron. If the next-to-leading order corrections are taken into account,
the renormalization/factorization scale dependence of the theoretical
prediction for the cross section is reduced considerably. The higher order
corrections increase the production cross section at the Tevatron by
about a factor two if we compare the next-to-leading order
prediction at a scale near the squark mass with the lowest order
prediction for which, in the experimental analyses, the scale was
identified with the invariant energy of the parton subprocess. This
results in a rise of the experimental lower bound on the squark mass
from the Tevatron by about $20$ GeV.

\pagebreak

The search for Higgs bosons and supersymmetric particles rank, {\it sub
  specie aeternitatis}, among
the most important experimental endeavors of high energy physics \cite{susy}.
Both areas are closely related through the hierarchy problem. In fact,
the idea of a supersymmetric extension of the standard model is
qualitatively  supported, though not proved, by the successful
prediction of the electroweak mixing angle. The colored
particles, i.e.~squarks and gluinos, of supersymmetric theories can be
searched for most
efficiently at the Tevatron $p\bar p$ collider and, in the
future, at the LHC up to very high mass values.
Lower bounds on the squark and gluino masses have been set by both
the Tevatron experiments CDF and D0. At the 90\% CL, the masses
are larger than about $148$ GeV for gluino masses below $400$
GeV\cite{massold,massnew}.

The evaluation of the experimental data has been based so far on the
lowest-order (LO) production cross sections \cite{born1,born2,born3,isajet},
Figs.~\ref{fig:feynman}(a) and (e).
In order to improve the theoretical predictions for the $p\bar p$
cross sections, the QCD corrections have
to be evaluated in next-to-leading order (NLO). The
theoretical analysis is straightforward but very tedious due to
the large variety of production channels and final states. We
will report in this letter, as a first indispensable step, on the
production of squark-antisquark pairs:
\begin{equation}
p\bar p \rightarrow \tilde{q} \bar{\tilde{q}} + X \quad\mbox{for}\quad
\tilde{q}\neq\tilde{t}
\end{equation}
The $\tilde q_L/\tilde q_R$ splitting, particularly
important for stop
particles, as well as the analysis of gluino final states
\cite{gluino} will be discussed in a more comprehensive report at a
later time. For the sake of
simplicity we have
taken all squark states mass degenerate, i.e.~the $n_{\tilde{q}}=5$
squark states in the final state as well as the stop particles, which
appear in internal loops. The only free parameters are
therefore the masses of the squarks and gluinos,
$m_{\tilde{q}}$ and $m_{\tilde{g}}$, respectively. The top mass is
fixed to $m_t = 174$ GeV \cite{topCDF}.

The next-to-leading order QCD calculation has been performed in the
Feynman gauge and the singularities have been isolated by means of
dimensional regularization. The masses have been renormalized in the
on-shell scheme. The massive particles are decoupled smoothly for
momenta smaller than their masses within the modified
$\overline{MS}$ scheme \cite{msbarext}. When removing the
infrared divergences, a cut-off $\Delta$ has been
introduced for the invariant mass of the squark-gluon
system in the final state, which separates soft from hard gluon
radiation \cite{top}. If both contributions are added, any
$\Delta$
dependence disappears from the total cross section for $\Delta\to
0$. The remaining collinear mass singularities can be
absorbed into the renormalization of the parton densities \cite{altar},
carried out in the $\overline{MS}$ factorization scheme. The GRV
parametrizations of the parton densities \cite{GRV} have been
adopted, which allow for proper LO
and NLO comparisons; CTEQ2 \cite{CTEQ} and MRSH \cite{MRS} parametrizations
have also been used for assessing the uncertainties from
the parton densities.

If the squarks are lighter than the gluinos, squarks can also be
decay products of on-shell gluinos, $\tilde g \rightarrow \tilde q \bar q$
etc.. Since we focus in the present analysis on the evaluation of the QCD
corrections, we shall restrict ourselves to irreducible final
states in which squarks do not evolve from on-shell gluinos; after
gluino final states are included explicitly, this
technical assumption will dissolve {\it eo ipso}. For the wedge
$m_{\tilde{q}} > m_{\tilde{g}}$ we disregard, in the same sense as above,
the decay of the squarks to gluinos.

Since quark-quark
pair initial states give negligible contributions to the generation
of high invariant mass states in $p\bar p$ collisions, the cross
section is built up primarily by quark-antiquark annihilation and, to
a lesser extent, by gluon fusion.

\paragraph*{Quark-antiquark initial states.}
The subprocesses $q\bar q\rightarrow\tilde q\bar{\tilde q}$ are
described to leading order, Fig.~\ref{fig:feynman}$(a_1)$ and $(a_2)$,
by mixtures of
amplitudes appropriate to unlike/like flavors and helicities.
Typical standard QCD and supersymmetric vertex
corrections are displayed in Figs.~\ref{fig:feynman}$(b_1)$ and
$(b_2)$. Ordinary
gluon radiation and a diagram related to the
renormalization of the parton
densities are exemplified in Figs.~\ref{fig:feynman}$(c)$ and $(d)$. Note that
$\tilde q\bar{\tilde q}$ pairs can be generated in $qg$
collisions only in next-to-leading order.

The diagrams have been evaluated analytically to obtain the double
differential cross sections $d\hat{\sigma}_{ij}/d\hat{t}d\hat{u}$ at the
parton level; $i,j$ are the parton indices $g,q,\bar q$ and
$\hat t$,$\hat u$ are the usual Mandelstam momentum
transfer variables. The total cross sections $\hat{\sigma}_{ij}$ may be
expressed in terms of scaling functions $f_{ij}$,

\begin{equation}
  \hat{\sigma}_{ij} = \frac{\alpha_s^2(Q^2)}{m_{\tilde
      q}^2}\left\{f_{ij}^{(0)}(\eta;r) +
  4\pi\alpha_s(Q^2)\left[f_{ij}^{(1)}(\eta;r,r_t)
    + \bar{f}_{ij}^{(1)}(\eta;r)\log\left(\frac{Q^2}{m_{\tilde
        q}^2}\right)\right]\right\}
\label{cross}
\end{equation}
They depend on the invariant parton energy $\sqrt{\hat s}$ through
$\eta = \hat s/4m_{\tilde q}^2 -1$, and on the ratios of the particle
masses $r =
m_{\tilde g}/m_{\tilde q}$, $r_t = m_t/m_{\tilde q}$. $\alpha_s$ is
the QCD coupling constant. Renormalization and
factorization scales are identified for the sake of simplicity,
$\mu_R = \mu_F = Q$. $f_{ij}^{(0)}$ denotes the lowest-order
contributions. While $f_{qg}^{(0)}$, the scaling function associated with
$qg$ collisions, is zero for $\tilde
q\bar{\tilde{q}}$ production, a compact expression \cite{born2} can be
found for quark-antiquark initial states,
\begin{eqnarray}
f_{q\bar q}^{(0)}(\eta;r) & = & \frac{\pi\beta}{108(1+\eta)}
\left\{
\left[\frac{1+r^2}{\beta(1+\eta)}+2\beta\right]6L_r -24 +
\frac{48r^2(1+\eta)}{(1+r^2)^2+4\beta^2 r^2 (1+\eta)}
\right. \nonumber \\
& & \left. +\,4n_{\tilde q}\beta^2
+2\beta\left[-\frac{L_r}{1+\eta}\left(r^2
+\frac{(1+r^2)^2}{4\beta^2(1+\eta)}\right) +2\beta
+\frac{1+r^2}{\beta(1+\eta)}\right]\right\}
\end{eqnarray}
\begin{eqnarray}
f_{q'\bar q}^{(0)}(\eta;r) &=& \frac{\pi\beta}{108(1+\eta)}
\left\{
\left[\frac{1+r^2}{\beta(1+\eta)}+2\beta\right]6L_r -24 +
\frac{48r^2(1+\eta)}{(1+r^2)^2+4\beta^2 r^2 (1+\eta)}\right\}
\end{eqnarray}
where $L_r = \log(x_+/x_-)$ with $x_\pm =1 +r^2 \pm 2\beta/(1\mp
\beta)$ and $\beta = (1 - 4m_{\tilde q}^2/\hat{s})^{1/2}$.
The scaling functions $f_{ij}^{(1)}$ and $\bar{f}_{ij}^{(1)}$,
describing the
next-to-leading order corrections, are displayed
in Fig.~\ref{fig:parton}. The scaling functions $f_{ij}^{(1)}$ are split into
the
"virtual + soft" part (V+S) and the "hard" part (H) into which the
infrared $\log^j\Delta$ ($j=1,2$) singularities of the (V+S)
contribution are absorbed
so that these functions are insensitive to the choice of $\Delta\to 0$.
Near the threshold, which is the kinematically most important region
in $p\bar p$ collisions, the Sommerfeld rescattering contribution
due to the exchange of Coulomb gluons between the slowly moving
$\tilde q\bar{\tilde q}$ pairs, leads to a singularity $\sim \pi
\alpha_s/\beta$ which neutralizes the phase space suppression near
threshold,
\begin{equation}
  f_{q\bar q}^{(1)thr}(\eta;r) = f_{q\bar q}^{(0)thr}(\eta;r)
  \left\{\frac{7}{48 \beta}
  +\frac{2}{3 \pi^2}\log^2(8\beta^2)
  -\frac{11}{4\pi^2}\log(8\beta^2)\right\}
\end{equation}
with
\begin{displaymath}
  f_{q\bar q}^{(0)thr}(\eta;r) =\frac{4\pi\beta r^2}{9(1+r^2)^2}
\end{displaymath}
The same relations hold for unlike flavors near threshold. The
$\log^2\beta$ terms, generated by initial state gluon radiation near
the threshold, can be exponentiated \cite{mueller}.
The plateaus for large parton energies are due to
flavor-excitation and gluon-splitting mechanisms
[cf.~Fig.~\ref{fig:feynman}($d$)]. The t- and u-channel
exchanges of gluons lead to an asymptotically constant
cross section, the scale of which is set by the squark mass,
i.e.~$\hat\sigma \sim \alpha_s^3/m_{\tilde q}^2$, to be contrasted
with the scaling
behavior $\hat\sigma \sim \alpha_s^2/s$ of the cross section to lowest
order. The values of the scaling functions in the asymptotic plateau
region can be calculated analytically, $f_{qg,H}^{(1)} \to 2159/(19440\pi)$
and $\bar{f}_{qg}^{(1)} \to -11/(324\pi)$.

\paragraph*{Gluon-gluon initial states.}
To lowest-order the diagrams contributing to the subprocess
$gg\rightarrow \tilde q\bar{\tilde q}$ are the well-known diagrams
from quark-pair production
[e.g.~Fig.~\ref{fig:feynman}$(e_1)$] supplemented by the seagull term
Fig.~\ref{fig:feynman}$(e_2)$ for scalar squarks.
Carrying out the next-to-leading order program, we
find, in the same notation as above \cite{born2},

\begin{equation}
  f_{gg}^{(0)}(\eta;r) = \frac{n_{\tilde q}\pi\beta
    }{96(1+\eta)^2}\left\{ \frac{41}{2} +5\eta
  +\left(\frac{8}{\beta} +
  \frac{1}{2\beta(1+\eta)}\right)
  \log\left(\frac{1-\beta}{1+\beta}\right) \right\}
\end{equation}
The $gg$ scaling functions are displayed in Fig.~\ref{fig:parton}(d) in leading
and
next-to-leading order.
Near the threshold, the Sommerfeld enhancement is observed again in
next-to-leading order,
\begin{equation}
  f_{gg}^{(1)thr}(\eta) = f_{gg}^{(0)thr}(\eta) \left\{
  \frac{11}{336\beta}
  +\frac{3}{2\pi^2}\log^2(8\beta^2)
  -\frac{183}{28\pi^2}\log(8\beta^2)\right\}
\end{equation}
with
\begin{displaymath}
  f_{gg}^{(0)thr}(\eta) = \frac{7n_{\tilde
    q}\pi\beta}{192}.
\end{displaymath}
The high-energy plateaus are generated as before by flavor-excitation and
gluon-splitting mechanisms, $f_{gg,H}^{(1)} \to 2159/(4320\pi)$
and $\bar{f}_{gg}^{(1)} \to -11/(72\pi)$.

While for $\alpha_s \sim 0.1$ the higher-order corrections are
suppressed significantly with respect to the Born term for quark
initial states, the large color charge of gluons leads to corrections of the
order of the Born term for the $gg$ initiated subprocess.

After these introductory remarks we present our final results in
Fig.~\ref{fig:limit} and Table \ref{tab:kfac}. The
cross sections of the various subprocesses have been convoluted with
the parton densities in the GRV \cite{GRV}, CTEQ2 \cite{CTEQ} and MRSH
\cite{MRS} parametrizations. The following
conclusions can be drawn from the analysis of the figures and the table.

(i) It is obvious from Fig.~3(a) that the theoretical
predictions for the $p\bar p \rightarrow \tilde q\bar{\tilde q}$
production process are improved considerably by taking into account
the next-to-leading order QCD corrections. While the
dependence on the renormalization/factorization scale $Q$ is quite
steep and monotonic in leading order, the $Q$ dependence is
significantly reduced in next-to-leading order
for reasonable variations of the scale, running even through a broad
maximum near $Q \sim m_{\tilde q}/3$. Since the cross section is
built up mainly by the quark channels [$\simgt 85\%$] and thus based on
well-measured parton densities, the variation between different
parton parametrizations is negligibly small.

(ii) The $K$ factors, defined as the ratio
$K=\sigma_{NLO}/\sigma_{LO}$ [with all quantities in the numerator and
denominator calculated consistently in NLO and LO, respectively],
depend only mildly on the squark and gluino masses. Experimental mass
bounds can
therefore be corrected easily for higher-order QCD effects. A sample of
$K$ factors is collected in Table \ref{tab:kfac} for various choices of the
renormalization/factorization scale parameter $Q$.

(iii) Finally in Fig.~3(b) we illustrate the impact of the QCD
corrections on the experimental lower bounds of the squark masses. We
compare the lowest-order cross section at $Q = \sqrt{\hat s}$, the
scale adopted in experimental analyses, with the
next-to-leading order prediction at $Q = m_{\tilde q}$
\cite{coll}. The NLO cross section is significantly larger at this
theoretically reasonable scale than the LO cross section at the scale
$\sqrt{\hat s}$. Taken at face value, this increases the bound on the
squark mass by about $20$ GeV. While no
precise value of $Q$ can be defined {\it a priori}, it is clear
nevertheless
that the experimental bounds derived from Tevatron data
\cite{massold,massnew} are {\it very} conservative and that the true bounds
are likely to be higher by as much as $\sim 20$ GeV.

We thank S. Lammel for useful discussions on the Tevatron squark and
gluino mass limits and their theoretical basis.

\pagebreak
\begin{table}
  \begin{center}
    \leavevmode
    \begin{tabular}{|c|c|c|c|c|}
      \hline
      $m_{\tilde q}$ (GeV) & $m_{\tilde g}$ (GeV) & $Q= m_{\tilde
        q}/3$ & $Q= m_{\tilde q}$ & $Q= 2 m_{\tilde q}$ \\
      \hline
      150 & 200 & 0.75 & 1.15 & 1.37 \\
      150 & 400 & 0.70 & 1.11 & 1.34 \\
      200 & 200 & 0.74 & 1.14 & 1.36 \\
      200 & 400 & 0.72 & 1.12 & 1.34 \\
      250 & 200 & 0.76 & 1.15 & 1.38 \\
      250 & 400 & 0.75 & 1.15 & 1.37 \\
      400 & 200 & 0.81 & 1.18 & 1.39 \\
      400 & 400 & 0.78 & 1.15 & 1.37 \\
      \hline
    \end{tabular}
  \end{center}
  \caption{$K$ factors for a set of $\tilde q, \tilde g$ masses and a
    range of renormalization/factorization scales $Q$ at the Tevatron
    energy $\protect\sqrt{s} =1.8$ TeV; GRV parton densities
    \protect\cite{GRV}.}
  \label{tab:kfac}
\end{table}

\vspace*{1.5cm}

\section*{Figures}
\begin{figure}[h]
  \caption{Generic diagrams for squark-antisquark pair production:
    $(a_1)$, $(a_2)$ basic Born-level $q\bar q$ diagrams; $(b_1)$ and $(b_2)$
    vertex corrections due to gluon and gluino exchange; $(c)$
    gluon emission; $(d)$ gluon-quark process;
    $(e_1)$, $(e_2)$ gluon fusion.}
  \label{fig:feynman}
\end{figure}
\begin{figure}[h]
  \caption{The scaling functions for squark-antisquark pair production
    in (a) $q\bar q$, (b) $q'\bar q$, (c) $qg$ and (d) $gg$
    collisions. The notation follows eq.(\protect\ref{cross}) with
    $\eta = \hat{s}/4m_{\tilde q}^2 -1$; V+S denotes the
    sum of the virtual and soft corrections, H the contribution of
    hard gluon emission. Mass parameters:
    $m_{\tilde q}= 250$ GeV and $m_{\tilde g} = 200$ GeV.}
  \label{fig:parton}
\end{figure}
\begin{figure}[h]
  \caption{Total cross section for the irreducible production of
    squark-antisquark pairs $p\bar p \to \tilde q
    \bar{\tilde q} X$ at the Tevatron energy $\protect\sqrt{s} = 1.8$
    TeV. (a)
   Dependence on the renormalization/factorization scale $Q$ for the
 leading order (LO) and the next-to-leading order (NLO) predictions,
 and sensitivity to different parton densities;
mass parameters as in Fig.2; (b) Dependence of the cross section
on the squark mass for $m_{\tilde g} = 200$ GeV; GRV parton
densities for NLO, and EHLQ parton densities for LO used in the
experimental analysis.
\protect\cite{massnew}. Upper full line
of the NLO prediction corresponds to the renormalization/factorization
scale $Q/m_{\tilde q} = 1/3$, middle line $=1$ and lower line $=2$.}
  \label{fig:limit}
\end{figure}


\vspace{2cm}
\begin{thebibliography}{xx}

\bibitem{susy}  H. P. Nilles, Phys. Rep. {\bf 110}, 1 (1984);
  H. E. Haber and G. L. Kane, Phys. Rep. {\bf 117}, 75 (1985).
\bibitem{massold} CDF Collaboration, F.~Abe {\it et al.},
  Phys. Rev. Lett. {\bf 69}, 3439 (1992).
\bibitem{massnew} S.~Hagopian,  Proceedings
  of the XXVII Int. Conf. on High Energy Physics, Glasgow (1994) and
  FERMILAB Conf 94/ 331-E (1994); D. R. Claes, FERMILAB Conf 94/ 290-E
  (1994);
  S.~Lammel, talk at the DESY Theory Workshop on Supersymmetry, Hamburg
  (1994).
\bibitem{born1} G. L. Kane and J. P. Leveille, Phys. Lett. B {\bf 112},
  227 (1982).
\bibitem{born2} P. R. Harrison and C. H. Llewellyn Smith,
  Nucl. Phys. {\bf B213}, 223 (1983); {\bf B223}, 542(E) (1983).
\bibitem{born3} E. Reya and D. P. Roy, Phys. Rev. D {\bf 32}, 645
  (1985); H. Baer and X. Tata, Phys. Lett. B {\bf 160}, 159 (1985).
\bibitem{isajet} H. Baer, F. E. Paige, S. D. Protopopescu and X. Tata,
  Report on ISAJET 7.0/ ISASUSY 1.0, FSU-HEP 930329 and UH-511-764-93.
\bibitem{gluino} The top-quark analog of the QCD corrections to
  gluino-pair production in $gg$ fusion, with squark loops left out
  however, has been presented in Ref. \cite{top}.
\bibitem{top} W. Beenakker, H. Kuijf, W. L. van Neerven and J. Smith,
  Phys. Rev. D {\bf 40}, 54 (1989).
\bibitem{topCDF} CDF Collaboration, F. Abe {\it et al.},
  Phys. Rev. Lett. {\bf 73}, 225 (1994); Phys. Rev. D {\bf 50}, 2966
  (1994).
\bibitem{msbarext} J. Collins, F. Wilczek and A. Zee,
  Phys. Rev. D {\bf 18}, 242 (1978);  W. J. Marciano, Phys. Rev. D {\bf
    29}, 580 (1984); P. Nason, S. Dawson and R. K. Ellis,
  Nucl. Phys. {\bf B303}, 607 (1988).
\bibitem{altar} G. Altarelli, R. K. Ellis and G. Martinelli,
  Nucl. Phys. {\bf B157}, 461 (1979); W. Furmanski and R. Petronzio,
  Z. Phys. C {\bf 11}, 293 (1982).
\bibitem{GRV} M. Gl\"uck, E. Reya and A. Vogt, Z. Phys. C {\bf 53}, 127 (1992).
\bibitem{CTEQ} J. Botts, J.G. Morfin, J.F. Owens, J. Qiu, W.-K. Tung
  and H. Weerts, CTEQ Collaboration, Phys. Lett. B {\bf 304}, 159
  (1993); J. Botts, H.L. Lai, J.G. Morfin, J.F. Owens, J. Qiu and
  W.-K. Tung, CTEQ2 Collaboration, MSUHEP-93/28 (1993).
\bibitem{MRS} A.D. Martin, W.J. Stirling and R.G. Roberts,
  Proceedings, Workshop on Quantum Field Theoretical Aspects of High
  Energy Physics, B. Geyer and E.-M. Ilgenfritz (eds.), ZHS
  Univ. Leipzig (1993).
\bibitem{mueller} A. H. Mueller and P. Nason, Phys. Lett. B {\bf 156},
  226 (1985); Nucl. Phys. {\bf B266}, 256 (1986).
\bibitem{coll} The scale $\protect\sqrt{\hat{s}}$ cannot be used in NLO; see
  J. C. Collins, D. E. Soper and G. Sterman in Perturbative Quantum
  Chromodynamics (ed. A. H. Mueller), World Scientific, Singapore (1989).
\end{thebibliography}
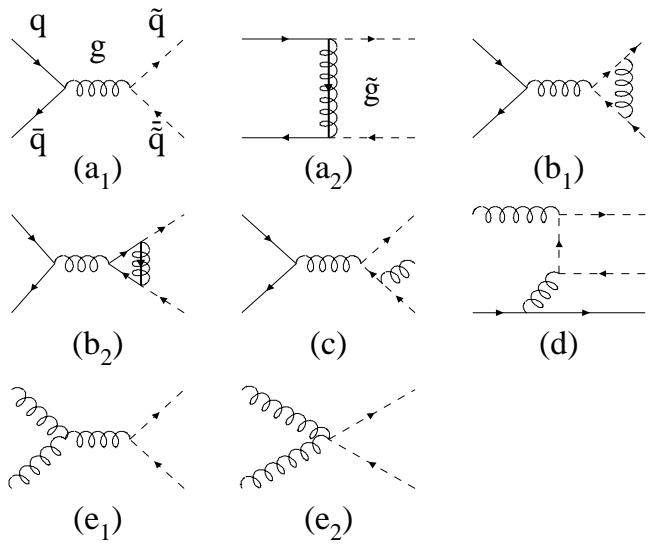
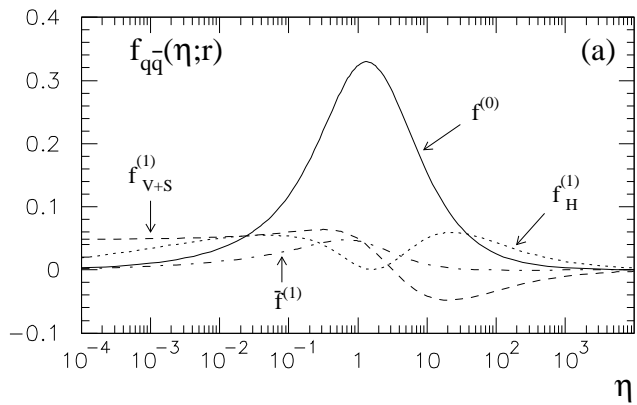
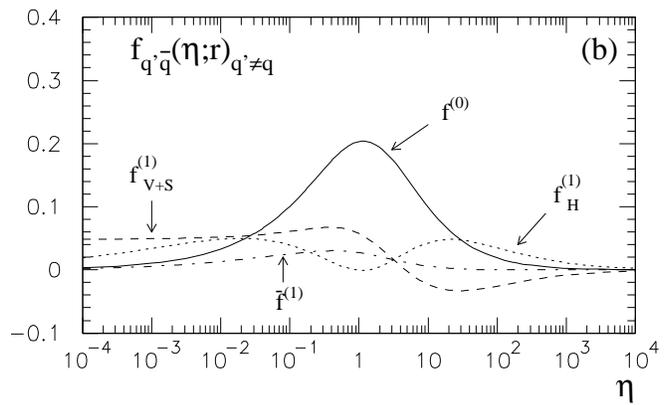
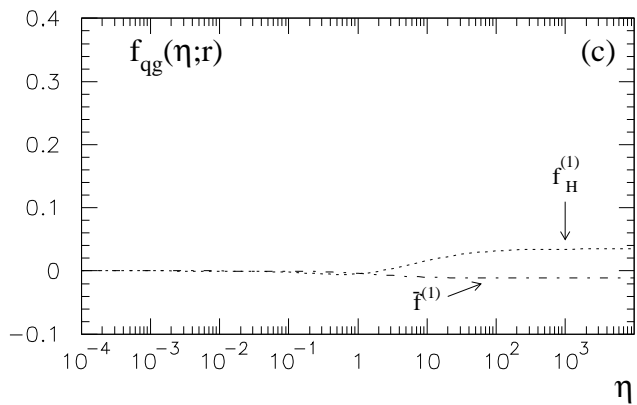
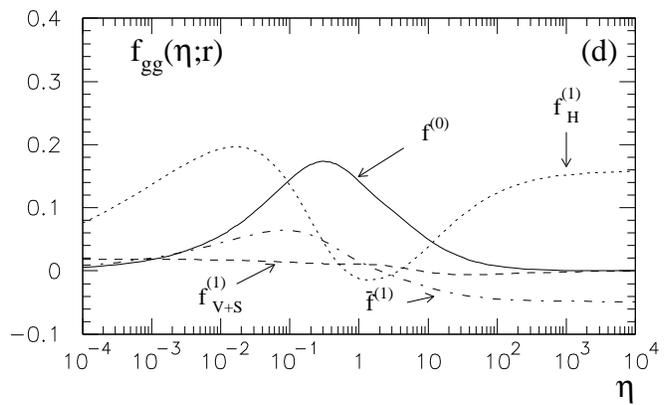
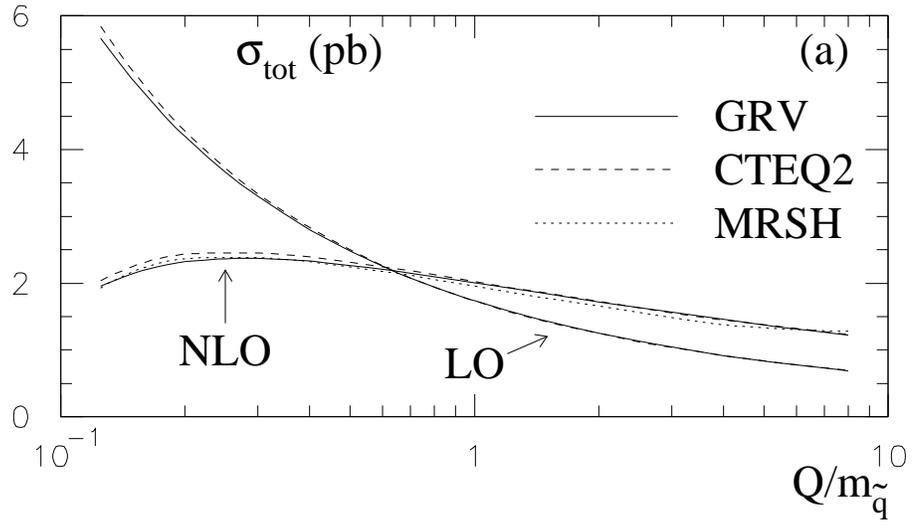
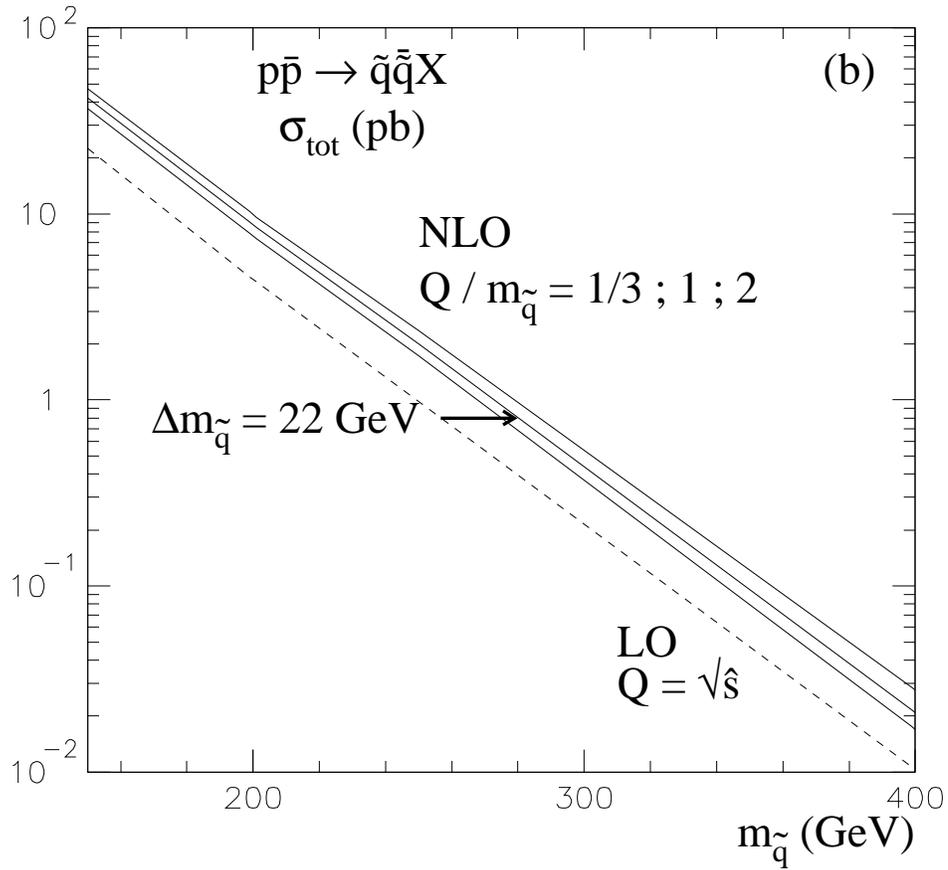
\end{document}